\documentclass[letterpaper]{article} % DO NOT CHANGE THIS
\usepackage{aaai25}  % DO NOT CHANGE THIS
\usepackage{times}  % DO NOT CHANGE THIS
\usepackage{helvet}  % DO NOT CHANGE THIS
\usepackage{courier}  % DO NOT CHANGE THIS
\usepackage[hyphens]{url}  % DO NOT CHANGE THIS
\usepackage{graphicx} % DO NOT CHANGE THIS
\usepackage{natbib}  % DO NOT CHANGE THIS AND DO NOT ADD ANY OPTIONS TO IT
\usepackage{caption} % DO NOT CHANGE THIS AND DO NOT ADD ANY OPTIONS TO IT
\usepackage{algorithm}
\usepackage{algorithmic}
\usepackage{amsmath}
\usepackage{booktabs}
\usepackage{subcaption}
\usepackage{multirow}
\usepackage{makecell}
\usepackage{xcolor}
\usepackage{colortbl}
\definecolor{zhz_gray}{rgb}{0.8,0.8,0.8}

\newcommand{\nop}[1]{}

\usepackage{newfloat}
\usepackage{listings}
\DeclareCaptionStyle{ruled}{labelfont=normalfont,labelsep=colon,strut=off} % DO NOT CHANGE THIS
\floatstyle{ruled}
\newfloat{listing}{tb}{lst}{}
\floatname{listing}{Listing}
\title{Revolutionizing Encrypted Traffic Classification with MH-Net: A Multi-View Heterogeneous Graph Model}
\author{
    Haozhen Zhang\textsuperscript{\rm 1,\rm 2,\rm 3}\equalcontrib,
    Haodong Yue\textsuperscript{\rm 1,\rm 2,\rm 3}\equalcontrib,
    Xi Xiao\textsuperscript{\rm 1,\rm 2,\rm 3}\thanks{Corresponding author.},
    Le Yu\textsuperscript{\rm 4},
    Qing Li\textsuperscript{\rm 2},
    Zhen Ling\textsuperscript{\rm 5},
    Ye Zhang\textsuperscript{\rm 6}
}
\affiliations{
    \textsuperscript{\rm 1}Tsinghua Shenzhen International Graduate School, Tsinghua University, Shenzhen, China\\
    \textsuperscript{\rm 2}Peng Cheng Laboratory, Shenzhen, China\\
    \textsuperscript{\rm 3}Key Laboratory of Cyberspace Security, Ministry of Education of China, Zhengzhou, China\\
    \textsuperscript{\rm 4}Nanjing University of Posts and Telecommunications, Nanjing, China\\
    \textsuperscript{\rm 5}Southeast University, Nanjing, China\\
    \textsuperscript{\rm 6}National University of Singapore, Singapore\\
    zhang-hz21@tsinghua.org.cn, yhd24@mails.tsinghua.edu.cn, xiaox@sz.tsinghua.edu.cn, yulele08@njupt.edu.cn, liq@pcl.ac.cn, zhenling@seu.edu.cn, e1124294@u.nus.edu
}

\usepackage{bibentry}
\begin{document}

\maketitle

\begin{abstract}

With the growing significance of network security, the classification of encrypted traffic has emerged as an urgent challenge. Traditional byte-based traffic analysis methods are constrained by the rigid granularity of information and fail to fully exploit the diverse correlations between bytes. To address these limitations, this paper introduces MH-Net, a novel approach for classifying network traffic that leverages multi-view heterogeneous traffic graphs to model the intricate relationships between traffic bytes. The essence of MH-Net lies in aggregating varying numbers of traffic bits into multiple types of traffic units, thereby constructing multi-view traffic graphs with diverse information granularities. By accounting for different types of byte correlations, such as header-payload relationships, MH-Net further endows the traffic graph with heterogeneity, significantly enhancing model performance. Notably, we employ contrastive learning in a multi-task manner to strengthen the robustness of the learned traffic unit representations.
Experiments conducted on the ISCX and CIC-IoT datasets for both the packet-level and flow-level traffic classification tasks demonstrate that MH-Net achieves the best overall performance compared to dozens of SOTA methods.

\end{abstract}

\begin{links}
    \link{Code}{https://github.com/ViktorAxelsen/MH-Net}
\end{links}

\section{Introduction}
\label{sec:intro}

As advancements in computer network technology continue and various devices connect to the Internet, user privacy becomes increasingly vulnerable to malicious attacks. 
While encryption technologies like VPNs and Tor~\cite{VPNTor} offer protection to users~\cite{Encryption, xiao2024comprehensive}, they can paradoxically serve as tools for attackers to conceal their identities. Traditional data packet inspection (DPI) methods have lost effectiveness against encrypted traffic~\cite{EncrySurvey}. Designing a universally effective method to classify an attacker's network activities (e.g., website browsing or application usage) from encrypted traffic remains a formidable challenge.

In the past few years, many methods have been proposed to enhance the capability of encrypted traffic classification techniques. 
Among them, statistic-based methods~\cite{AppScanner, KFP, FlowPrint, CUMUL, ETC-PS} generally rely on hand-crafted traffic statistical features and then leverage a traditional machine learning model for classification. 
However, they require heavy feature engineering and are susceptible to unreliable flows~\cite{TFE-GNN}. 
With the burgeoning of representation learning~\cite{CLSurvey}, some methods also use deep learning models to conduct traffic classification, such as pre-trained language models~\cite{ETBERT, PacRep}, neural networks~\cite{FSNet, TFE-GNN, YaTC}, etc. 
Although these methods demonstrate competitive performance, they fail to sufficiently uncover the fine-grained correlations between traffic bytes, which can be attributed to the following two disadvantages. (1) \textbf{Rigid information granularity constrained by bytes}. Most existing methods treat a byte as an indivisible unit by default, which ignores the diverse granularity of information contained in traffic data. To give a practical example, a Chinese character is represented by two bytes, while an English character is represented by only one byte, indicating that traffic data contains information of different granularities universally (note that the granularity is not necessarily bytes, it may also be bits). 
(2) \textbf{Lack of consideration of multiple correlation types between bytes}. Current methods mix the correlation of bytes at different positions in the byte sequence, which overlooks and fails to utilize the differences between diverse types of correlation (e.g., the correlation types between bytes in the header and bytes in the payload are different). 
Consequently, how to uncover and leverage the potential fine-grained correlations between traffic bytes to enhance traffic classification is a salient problem.

To address the above challenges, in this paper, we propose a novel model named \textbf{MH-Net}, which classifies network traffic using multi-view heterogeneous traffic graphs. 
In particular, MH-Net first aggregates different numbers of traffic bits into multiple types of traffic units (e.g., 4-bit units and 8-bit units), which facilitates the diversity of information granularity of traffic data. 
Since graphs have excellent capabilities in processing relational data, MH-Net further leverages point-wise mutual information (PMI) to convert diverse types of traffic unit sequences into multi-view traffic graphs. 
On top of this, considering the distinct functionality in different parts of the traffic unit sequence, MH-Net introduces three types of unit correlations, i.e., header-header, header-payload, and payload-payload unit correlations, and designs a heterogeneous traffic graph encoder for multi-view heterogeneous graph representation learning. 
To strengthen the robustness of traffic unit representations, we additionally perform contrastive learning in a multi-task manner on traffic graphs. 
To comprehensively evaluate MH-Net, we conduct both the packet-level and flow-level traffic classification tasks on the ISCX and CIC-IoT datasets. 
Experimental results show that MH-Net achieves competitive performance and ranks first among all the state-of-the-art methods. 
Further analysis of traffic units also reveals the potential trade-off between complementarity and interference between traffic units with different
information granularity.

To summarize, our main contributions include: 
\begin{itemize}
\item We propose a novel model named MH-Net, which aggregates different numbers of traffic bits into multiple types of traffic units to construct multi-view traffic graphs, enriching the diversity of information granularity while improving model performance. 
\item MH-Net further introduces three types of traffic unit correlations to model the heterogeneity of traffic graphs and employs a heterogeneous graph neural network for feature extraction. Furthermore, we conduct contrastive learning in a multi-task manner to enhance the robustness of traffic unit representations.
\item We conduct experiments on both the packet-level and flow-level traffic classification tasks using the ISCX and CIC-IoT datasets. The experimental results show that MH-Net achieves the overall best performance compared with dozens of baselines. 
\end{itemize}

\section{Related Work}
\label{sec:related_work}

\noindent \textbf{Flow-level Traffic Classification Methods.} 
Flow-level traffic classification methods aim to classify traffic flows, which can be summarized into three categories. 

$\bullet$ \noindent\textit{Statistical Feature Based Methods.} 
Many methods use statistical features to represent packet properties and utilize traditional machine learning models for classification. 
AppScanner~\cite{AppScanner} extracts features from traffic flows based on bidirectional flow characteristics, while CUMUL~\cite{CUMUL} uses cumulative packet length as its feature. 
ETC-PS~\cite{ETC-PS} strengthens packet length sequences by applying the path signature theory, and hierarchical clustering is also leveraged for feature extraction by Conti \emph{et al.}~\cite{Conti}. 

$\bullet$ \noindent\textit{Fingerprinting Matching Based Methods.} 
Fingerprinting denotes traffic characteristics and is also used in traffic identification. 
FlowPrint~\cite{FlowPrint} generates traffic fingerprints by creating correlation graphs that compute activity values between destination IPs. 
K-FP~\cite{KFP} uses the random forest to construct fingerprints and identifies unknown samples through k-nearest neighbor matching. 

$\bullet$ \noindent\textit{Deep Learning Based Methods.} 
Deep learning has demonstrated powerful learning abilities, and many traffic classification methods are based on it. 
RBRN~\cite{RBRN}, DF~\cite{DF}, and FS-Net~\cite{FSNet} all use statistical feature sequences (e.g., packet length sequences) as inputs for convolutional neural networks (CNNs) or recurrent neural networks (RNNs). 
Additionally, there are some methods using raw bytes as features. 
EBSNN~\cite{EBSNN} combines RNNs with the attention mechanism to process header and payload byte segments. 
ET-BERT~\cite{ETBERT} conducts pre-training tasks on large-scale traffic datasets to learn a powerful raw byte representation, which is time-consuming and expensive. 
Graph neural networks (GNNs) are another model that can be used for traffic classification tasks. 
GraphDApp~\cite{GraphDApp} builds traffic interaction graphs from traffic bursts and uses GNNs for representation learning. 
TFE-GNN~\cite{TFE-GNN} employs point-wise mutual information~\cite{PMI} to construct byte-level traffic graphs and designs a traffic graph encoder for feature extraction. 
YaTC~\cite{YaTC} features a masked autoencoder-based traffic transformer to enable efficient feature extraction while boosting performance.

\noindent \textbf{Packet-level Traffic Classification Methods.} In contrast, packet-level traffic classification methods identify diverse categories of each network packet. 
Securitas~\cite{Securitas} generates n-grams for raw bytes and utilizes Latent Dirichlet Allocation (LDA) to form protocol keywords as features, followed by SVM, C4.5 decision tree, or Bayes network for packet classification. 
2D-CNN~\cite{2DCNN} and 3D-CNN~\cite{3DCNN} treat packet bytes as pixel values and convert them into images, which are further fed into 2D-CNNs and 3D-CNNs for packet classification. 
DP~\cite{DeepPacket} leverages CNNs and autoencoders to extract byte features. 
BLJAN~\cite{BLJAN} explores the correlation between packet bytes and their labels and encodes them into a joint embedding space to classify packets. 
EBSNN~\cite{EBSNN} and ET-BERT~\cite{ETBERT} can also perform packet classification. But, they still require two independent training or fine-tuning to conduct flow-level and packet-level tasks, which are computationally expensive. 
PacRep~\cite{PacRep} leverages triplet loss~\cite{tripletloss} without data augmentation and jointly optimizes multiple packet-level tasks to learn a better packet representation. 

In summary, existing methods conduct traffic classification tasks without sufficiently considering the informative correlations contained in raw bytes, thus facing performance bottlenecks.

\begin{figure*}[htb]
	\centering
	\includegraphics[width=1.0\linewidth]{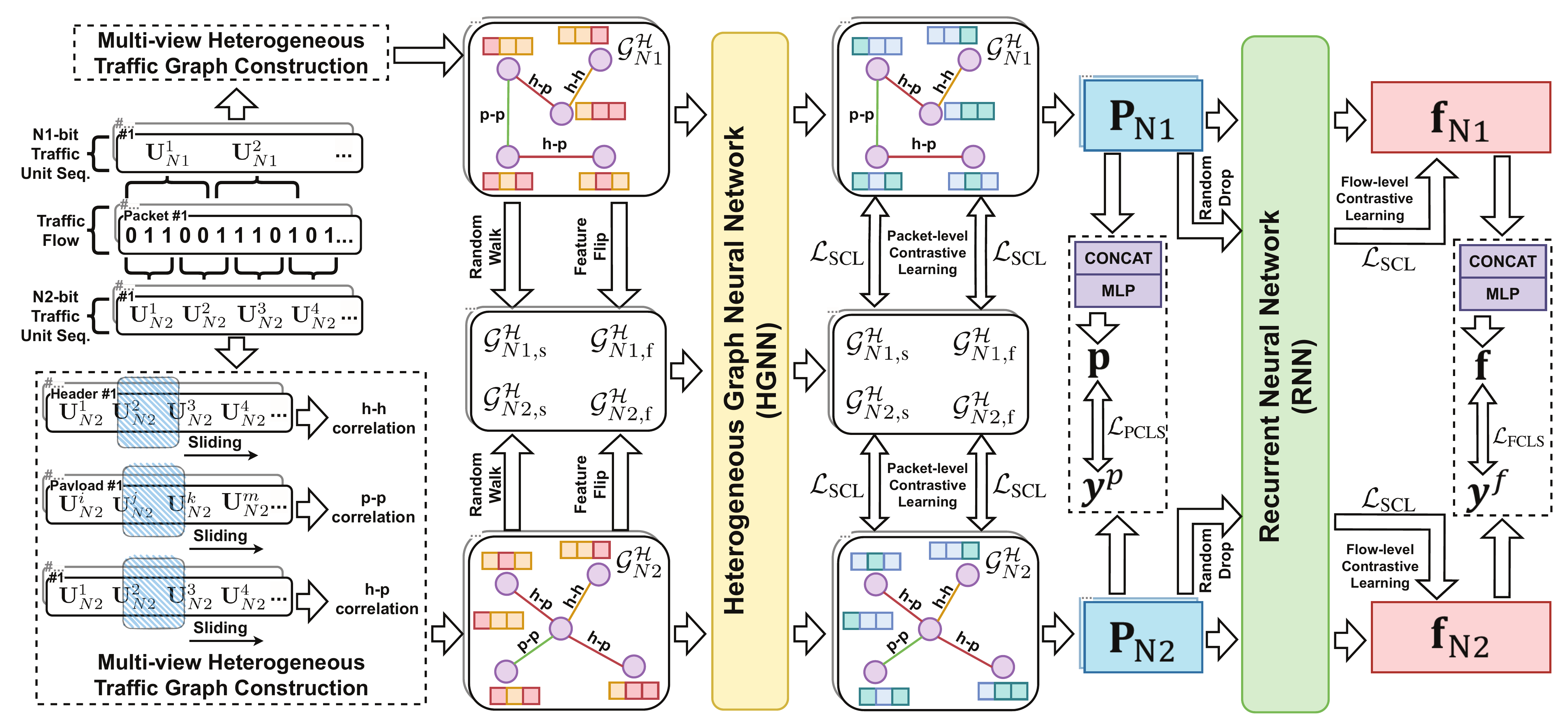}
        \caption{MH-Net Model Architecture. }
	\label{model_figure}
\end{figure*}

\section{Methodology}
\label{sec:method}

In this section, we will introduce the components of MH-Net, which include multi-view traffic graph construction, heterogeneous traffic graph representation learning, and multi-task training of MH-Net. The MH-Net model architecture is shown in Figure 1. 

\subsection{Multi-View Traffic Graph Construction}

In this section, we first describe the rationale behind the utilization of traffic units and detail the construction of multi-view traffic graphs using traffic units. 

\subsubsection{Traffic Units with Diverse Granularity.} 
To uncover the diverse information granularity contained in raw bytes, we attempt to seek more beyond the byte itself and instead aggregate raw bits of different lengths into various traffic units. 
Traffic units with different bit numbers can ``interpret" or ``express" the transmitted data from distinct perspectives, thereby being able to mine potential highly discriminative features. 
Taking character encoding as an example, given binary data with 16 bits, it can be converted into a Chinese character using 16-bit traffic units or 2 English characters using 8-bit traffic units. 
Therefore, we can draw on various traffic units to perform multi-view feature extraction on the same traffic data to derive better traffic representations.

\subsubsection{Traffic Graph Construction.} 
Although we aggregate traffic data into traffic units, they are still scattered individuals, which is not conducive to mining the potential fine-grained correlations between a sequence of traffic units. 
Inspired by~\cite{TFE-GNN, PMI}, due to the natural advantages of graph structure in data correlation modeling, we further convert the traffic unit sequence into a traffic graph to make it an interconnected whole. 
For a sequence of N-bit traffic units, we leverage point-wise mutual information (PMI) to quantify the correlation between traffic units. 
PMI employs a sliding window on a traffic unit sequence and counts the number of times two units co-occur within the window. The co-occurring frequency of the two units and the frequency of each unit are used to compute PMI, which can be formulated as:
\begin{equation}
p(\mathbf{U}_N^i, \mathbf{U}_N^j)=\frac{\# W(\mathbf{U}_N^i, \mathbf{U}_N^j)}{\# W}, \quad
p(\mathbf{U}_N^i)=\frac{\# W(\mathbf{U}_N^i)}{\# W}
\end{equation}
\begin{equation}
\operatorname{PMI}(\mathbf{U}_N^i, \mathbf{U}_N^j)=\log \frac{p(\mathbf{U}_N^i, \mathbf{U}_N^j)}{p(\mathbf{U}_N^i) p(\mathbf{U}_N^j)}
\end{equation}
where $\# W$ is the total number of sliding windows, $\# W(\mathbf{U}_N^i)$ is the number of times unit $\mathbf{U}_N^i$ appears in the sliding window, and $\# W(\mathbf{U}_N^i, \mathbf{U}_N^j)$ denotes the number of times units $\mathbf{U}_N^i$ and $\mathbf{U}_N^j$ appear in the sliding window simultaneously. 
We connect two traffic units $\mathbf{U}_N^i$ and $\mathbf{U}_N^j$ if and only if $\operatorname{PMI}(\mathbf{U}_N^i, \mathbf{U}_N^j) > 0$ is satisfied, which indicates high unit correlation. 
For the traffic data in each packet, we select two types of traffic units, e.g., N1-bit and N2-bit traffic units, and apply the PMI algorithm on their sequences to construct the multi-view traffic graphs $\mathcal{G}_{N1}$ and $\mathcal{G}_{N2}$ for each packet, respectively. Note that the node features of the traffic graph are the value of each traffic unit.

\subsection{Heterogeneous Traffic Graph Representation Learning}

This section will present the further integration of heterogeneous correlations in traffic graphs. Then, a heterogeneous traffic graph encoder is elaborated to conduct heterogeneous traffic graph representation learning. 

\subsubsection{Heterogeneous Traffic Unit Correlations.} 
In the byte sequence of a packet, the header and payload have information heterogeneity due to their different functions (the former carries the metadata of the packet, and the latter carries the actual transmitted content). 
However, the homogeneous correlations between traffic units in $\mathcal{G}$ restrict the full utilization of the heterogeneity in the header and payload. 
Therefore, we argue that the correlations between traffic units should have multiple types according to their positions in the traffic unit sequence. 
Intuitively, we propose three types of traffic unit correlations, i.e., header-header (h-h), payload-payload (p-p), and header-payload (h-p) correlations, to alleviate the above issue. 
We then apply the PMI algorithm to the traffic unit sequences of the header, payload, and header + payload (i.e., the entire traffic unit sequence), respectively, to obtain three types of traffic edges, which are further integrated into a heterogeneous traffic graph $\mathcal{G}^{\mathcal{H}}$. 

In this way, we transform $\mathcal{G}_{N1}$ and $\mathcal{G}_{N2}$ into multi-view heterogeneous traffic graphs $\mathcal{G}^{\mathcal{H}}_{N1}$ and $\mathcal{G}^{\mathcal{H}}_{N2}$ whereby the correlations between traffic units becomes more fine-grained, enabling seamless heterogeneous fusion of the header and payload in a single traffic graph.

\subsubsection{Heterogeneous Traffic Graph Encoder.} 

So far, we have obtained traffic graphs $\mathcal{G}^{\mathcal{H}}_{N1}$ and $\mathcal{G}^{\mathcal{H}}_{N2}$ for each packet, which will be further fed into a heterogeneous traffic graph encoder for traffic representation learning. 
Specifically, the heterogeneous traffic graph encoder features a heterogeneous graph neural network (HGNN) to extract discriminative features of traffic graphs. 
HGNN employs GraphSAGE~\cite{GraphSAGE} as its basic backbone, and the model weights are not shared across edge types. 
Generally, the forward pass of HGNN in the $l$-th layer can be described as:
\begin{equation}
\mathbf{m}_v^{(l)} = \operatorname{MSG}^{(l)}\left(\left\{\mathbf{h}_u^{(l-1)}, u \in N(v)\right\} ; \theta_{h \text{-} h}^{l,m}, \theta_{p \text{-} p}^{l,m}, \theta_{h \text{-} p}^{l,m}\right)
\end{equation}
\begin{equation}
\mathbf{h}_v^{(l)}=\operatorname{AGG}^{(l)}\left(\mathbf{h}_v^{(l-1)}, \mathbf{m}_v^{(l)} ; \theta_{h \text{-} h}^{l,a}, \theta_{p \text{-} p}^{l,a}, \theta_{h \text{-} p}^{l,a}\right)
\end{equation}
where $\mathbf{h}_{v}^{(l)}$ are the embedding vectors of nodes $v$ in layer $l$, $\mathbf{m}_{v}^{(l)}$ is the computed neighbor message for node $v$ in layer $l$, and $N(v)$ is the neighbors of node $v$. 
$\mathrm{MSG}^{(l)}(\cdot)$ is a message computation function parameterized by $\theta_{h \text{-} h}^{l,m}, \theta_{p \text{-} p}^{l,m}, \theta_{h \text{-} p}^{l,m}$ and 
$\mathrm{AGG}^{(l)}(\cdot)$ is a message aggregation function parameterized by $\theta_{h \text{-} h}^{l,a}, \theta_{p \text{-} p}^{l,a}, \theta_{h \text{-} p}^{l,a}$ in layer $l$. 
Then, we perform the element-wise average operation on all the node embedding vectors in the last layer to obtain the final packet-level traffic representation $\mathbf{p}_{N1}$ and $\mathbf{p}_{N2}$ for $\mathcal{G}^{\mathcal{H}}_{N1}$ and $\mathcal{G}^{\mathcal{H}}_{N2}$, respectively, which can be simplified as:
\begin{equation}
\mathbf{p}_{N1} = \operatorname{HGNN}(\mathcal{G}^{\mathcal{H}}_{N1}), \quad
\mathbf{p}_{N2} = \operatorname{HGNN}(\mathcal{G}^{\mathcal{H}}_{N2})
\end{equation}
We can further obtain the flow-level traffic representation $\mathbf{f}_{N1}$ and $\mathbf{f}_{N2}$ using a recurrent neural network (RNN):
\begin{equation}
\mathbf{f}_{N1} = \operatorname{RNN}(\mathbf{p}_{N1}^1,...,\mathbf{p}_{N1}^L), \quad
\mathbf{f}_{N2} = \operatorname{RNN}(\mathbf{p}_{N2}^1,...,\mathbf{p}_{N2}^L)
\end{equation}
where $L$ is the length of a traffic flow.

\subsection{Multi-Task Training of MH-Net}

We aim to jointly train MH-Net in a multi-task manner for better optimization. The training objective of MH-Net mainly includes traffic classification and contrastive learning tasks, which will be presented in detail below. 

\subsubsection{Traffic Classification Tasks.} 

In this work, we conduct the flow-level and packet-level traffic classification tasks in MH-Net simultaneously. 
As shown in Figure 1, we utilize an unshared traffic classifier with a multilayer perceptron (MLP) to transform the flow-level and packet-level representations and calculate the corresponding traffic classification task loss, respectively: 
\begin{equation}
\mathbf{f} = \operatorname{CONCAT}(\mathbf{f}_{N1}, \mathbf{f}_{N2}), \quad
\mathbf{p} = \operatorname{CONCAT}(\mathbf{p}_{N1}, \mathbf{p}_{N2})
\end{equation}
\begin{equation}
\mathcal{L}_{\text{FCLS}} = \operatorname{CE}(\operatorname{MLP}(\mathbf{f}), \boldsymbol{y}^f), \quad
\mathcal{L}_{\text{PCLS}} = \operatorname{CE}(\operatorname{MLP}(\mathbf{p}), \boldsymbol{y}^p)
\end{equation}
where $\operatorname{CONCAT}(\cdot)$ denotes concatenation operation, $\operatorname{CE}(\cdot)$ is the cross entropy loss function, $\boldsymbol{y}^f$ is the flow label, and $\boldsymbol{y}^p$ is the packet label that is consistent with the flow label it belongs to.

\subsubsection{Contrastive Learning at Dual Levels.}

Inspired by the powerful representation learning capabilities of contrastive learning~\cite{CPC, MoCo, SimCLR}, which aims to learn semantic-invariant representations by contrasting positive and negative sample pairs derived from various data augmentation, we propose to utilize it to further enhance the multi-view packet-level and flow-level traffic representations in MH-Net. 
In particular, MH-Net features a supervised contrastive loss~\cite{SCL} to take better advantage of the data labels during training, which can be formulated as: 
\begin{equation}
\mathcal{L}_{\text{SCL}}(\mathbf{z})=\sum_{i \in I} \frac{-1}{|M(i)|} \sum_{m \in M(i)} \!\! \log \frac{\exp \left(\mathbf{z}_i \cdot \mathbf{z}_m / \tau\right)}{\sum_{k \in K(i)} \exp \left(\mathbf{z}_i \cdot \mathbf{z}_k / \tau\right)}
\end{equation}
where $\mathbf{z}$ refers to the embedding vector collection of two groups of augmented data samples from the same source. 
$i \in I$ is the index of an arbitrary augmented sample in $\mathbf{z}$, $K(i) \equiv I \backslash\{i\}$, and $M(i) \equiv\left\{m \in K(i): \boldsymbol{y}_m=\boldsymbol{y}_i\right\}$ is the indices of all positive samples of $i$, conditioned on the same data label. 
The symbol $\cdot$ represents the inner product operation and $\tau \in \mathcal{R}^{+}$ denoted the temperature parameter.

\textbf{Packet-level Contrastive Learning.} 
Since the packet-level traffic representation is derived based on the traffic graph, we need to generate an augmented graph for further contrastive learning~\cite{GraphCL, GCC}. 
In this work, we mainly adopt two augmentation approaches: graph structure and node feature augmentation. 
For graph structure augmentation, we leverage the random walk algorithm~\cite{random_walk}, which starts by randomly picking a node and iteratively performs a random traversal on its neighbors to obtain the augmented traffic graph $\mathcal{G}^{\mathcal{H}}_{N,\text{s}}$. 
As for node feature augmentation, we augment the traffic graph by flipping its node features~\cite{graphmae}, resulting in another augmented traffic graph $\mathcal{G}^{\mathcal{H}}_{N,\text{f}}$. 
After obtaining the embedding vector of $\mathcal{G}^{\mathcal{H}}_{N,\text{s}}$ and $\mathcal{G}^{\mathcal{H}}_{N,\text{f}}$, the packet-level contrastive loss can be formulated as:
\begin{equation}
\begin{aligned}
\mathcal{L}_{\text{PCL}} = \sum_{i}^{\{1,2\}} \mathcal{L}_{\text{SCL}}(\{\operatorname{HGNN}(\mathcal{G}^{\mathcal{H}}_{Ni,\text{s}}), \mathbf{p}_{Ni}\}) + \\ \mathcal{L}_{\text{SCL}}(\{\operatorname{HGNN}(\mathcal{G}^{\mathcal{H}}_{Ni,\text{f}}), \mathbf{p}_{Ni}\})
\end{aligned}
\end{equation}
Notably, we treat the embedding vector $\mathbf{p}_{N}$ of the original traffic graph $\mathcal{G}^{\mathcal{H}}_{N}$ as an ``anchor" without augmentation for training stability (with slight abuse of notation). By directly perturbing structures and features on the traffic graph, the semantic-invariant representations contained in traffic units (i.e., nodes) can be well-uncovered through contrastive learning, thus enhancing model performance.

\textbf{Flow-level Contrastive Learning.} 
We further conduct the flow-level contrastive learning task by randomly dropping packets from a traffic flow with a certain probability $P_{\text{PD}} \in [0, 1]$ and obtain an augmented traffic flow. 
Similarly, the flow-level contrastive loss can be formulated as:
\begin{equation}
\mathcal{L}_{\text{FCL}} = \sum_{i}^{\{1,2\}} \mathcal{L}_{\text{SCL}}(\{\operatorname{RNN}(\mathbf{p}_{Ni}^1 \odot \rho_1,...,\mathbf{p}_{Ni}^L  \odot \rho_L), \mathbf{f}_{Ni}\})
\end{equation}
where $\rho_i \in\{0,1\}$ is drawn from a Bernoulli distribution $\rho_i \sim \mathcal{B}(P_{\text{PD}})$, denoting whether to drop the packet. 
Such learning paradigms can also help the model capture the common characteristics of the traffic flow through augmentation, leading to a robust flow-level representation.

\subsubsection{Overall Training Objective.} 
In summary, we propose the overall end-to-end multi-task training objective of MH-Net as follows: 
\begin{equation}
\mathcal{L} = \mathcal{L}_{\text{PCLS}} + \mathcal{L}_{\text{FCLS}} + \alpha \mathcal{L}_{\text{PCL}} + \beta \mathcal{L}_{\text{FCL}}
\end{equation}
where $\alpha, \beta \in [0, 1]$ are the coefficients that control the contribution of the packet-level and flow-level contrastive tasks, respectively.

\section{Experiments}
\label{sec:exp}

\subsection{Experimental Settings}
\label{sec:exp_setttings}

\subsubsection{Dataset}

To thoroughly evaluate MH-Net on the packet-level and flow-level traffic classification tasks, we adopt the CIC-IoT~\cite{ciciot}, ISCX VPN-nonVPN~\cite{ISCX} and ISCX Tor-nonTor~\cite{ISCX-Tor} datasets. 
In the experiment, we conduct all experiments independently on these five datasets, i.e., CIC-IoT, ISCX-VPN, ISCX-NonVPN, ISCX-Tor, and ISCX-NonTor. 

Since our method performs both the flow-level and packet-level classification tasks, we first adopt stratified sampling to partition the flow-level training and testing dataset into 9:1 according to the number of traffic flows for all datasets. All packets in the flow-level training and testing datasets are directly used as the packet-level training and testing datasets, respectively. The category of each packet is consistent with the traffic flow to which it belongs.

\subsubsection{Implementation Details and Baselines}

Since the value range of an N-bit traffic unit is determined by the number of bits it contains (i.e., $2^N$), we utilize 4-bit and 8-bit traffic units to construct multi-view heterogeneous traffic graphs to achieve a trade-off between diversity and computational costs. 
The max flow length (i.e., the max packet number within a flow) is set to 15. 
The number of layers in HGNN is set to 4, and we initialize the RNN to LSTM in MH-Net by default. 
As for the random walk, we set the scale of subgraphs following~\cite{GCC}, and the restart probability is 0.8. 
We set the packet dropping ratio $P_{\text{PD}}$ to 0.6, and the temperature coefficient $\tau$ is 0.07. 
The objective coefficients $\alpha$ and $\beta$ are set to 1.0 and 0.5, respectively. 
We implement MH-Net and conduct all experiments with PyTorch and Deep Graph Library. The experimental results are reported as the mean over five runs on an NVIDIA RTX 3080.

As for evaluation metrics, we use Overall Accuracy (AC) and Macro F1-score (F1). We compare MH-Net with the flow-level and packet-level methods for a comprehensive comparison. The comparison baselines include \textbf{Flow-level Traffic Classification Methods} (i.e., AppScanner~\cite{AppScanner}, K-FP (K-Fingerprinting)~\cite{KFP}, CUMUL~\cite{CUMUL}, ETC-PS~\cite{ETC-PS}, FS-Net~\cite{FSNet}, DF~\cite{DF}, ET-BERT~\cite{ETBERT}, GraphDApp~\cite{GraphDApp}, TFE-GNN~\cite{TFE-GNN}, and YaTC~\cite{YaTC}) and \textbf{Packet-level Traffic Classification Methods} (i.e., Securitas~\cite{Securitas}, 2D-CNN~\cite{2DCNN}, 3D-CNN~\cite{3DCNN}, DeepPacket (DP)~\cite{DeepPacket}, BLJAN~\cite{BLJAN}, and EBSNN~\cite{EBSNN}).

For the rest of the experiment section, we will evaluate MH-Net from the following research questions: 

\noindent \textbf{RQ1}: How does MH-Net perform on the packet-level and flow-level classification tasks? 

\noindent \textbf{RQ2}: How much does each module of MH-Net contribute to the model performance? 

\noindent \textbf{RQ3}: How sensitive is MH-Net to hyper-parameters, and how does the choice and combination of traffic units affect model performance?

\subsection{Comparison Experiments (RQ1)}
\label{sec:exp_results}

% Public Datasets
\begin{table*}[t]
    \centering
  \begin{tabular}{lccccccccccc}
    \toprule
    \multirow{2}{*}{Model} & \multicolumn{2}{c}{CIC-IoT} & \multicolumn{2}{c}{ISCX-Tor} & \multicolumn{2}{c}{ISCX-nonTor} & \multicolumn{2}{c}{ISCX-VPN} & \multicolumn{2}{c}{ISCX-nonVPN} & \multirow{2}{*}{\makecell{Avg.\\Rank}} \\
    \cmidrule(r){2-3} \cmidrule(r){4-5} \cmidrule(r){6-7} \cmidrule(r){8-9} \cmidrule(r){10-11}
    \multicolumn{1}{c}{} & AC & F1 & AC & F1 & AC & F1 & AC & F1 & AC & F1 & \multicolumn{1}{c}{} \\
    \midrule
    % Results
    AppScanner
    & 0.9674 & 0.9620
    & 0.7543 & 0.6163
    & 0.9153 & 0.8273 
    & 0.8889 & 0.8722
    & 0.7576 & 0.7486
    & 6 \\
    K-FP
    & 0.9349 & 0.9254
    & 0.7771 & 0.6313
    & 0.8741 & 0.8167 
    & 0.8713 & 0.8747
    & 0.7551 & 0.7387
    & 8 \\
    CUMUL
    & 0.9153 & 0.9101
    & 0.6686 & 0.4997 
    & 0.8605 & 0.7627 
    & 0.7661 & 0.7644
    & 0.6187 & 0.5897
    & 9 \\
    ETC-PS
    & 0.9218 & 0.9122
    & 0.7486 & 0.6033
    & 0.9365 & 0.8486 
    & 0.8889 & 0.8851
    & 0.7273 & 0.7208
    & 7 \\
    \midrule
    FS-Net
    & \underline{0.9805} & \underline{0.9759}
    & 0.8286 & 0.7242 
    & 0.9278 & 0.8285 
    & 0.9298 & 0.9234
    & 0.7626 & 0.7555
    & 5 \\
    DF
    & 0.8664 & 0.8601
    & 0.6514 & 0.4719
    & 0.8568 & 0.7590 
    & 0.8012 & 0.7921
    & 0.6742 & 0.6701
    & 10 \\
    ET-BERT
    & 0.9565 & 0.9122
    & 0.9543 & 0.9397
    & 0.9029 & 0.8332 
    & 0.9532 & 0.9463
    & \textbf{0.9167} & \underline{0.9235}
    & 4 \\
    GraphDApp
    & 0.6808 & 0.7263
    & 0.4286 & 0.2281
    & 0.6936 & 0.5352 
    & 0.6491 & 0.5740
    & 0.4495 & 0.3614
    & 11 \\
    TFE-GNN
    & 0.9699 & 0.9666
    & \textbf{0.9886} & 0.9855
    & 0.9390 & 0.8507 
    & 0.9591 & 0.9536
    & 0.9040 & \textbf{0.9240}
    & 2 \\
    YaTC
    & 0.9397 & 0.9105
    & \underline{0.9868} & \underline{0.9869}
    & \textbf{0.9579} & \textbf{0.9522} 
    & \underline{0.9605} & \underline{0.9671}
    & 0.7546 & 0.7544
    & 3 \\
    \midrule
    \textbf{MH-Net}
    & \textbf{0.9900} & \textbf{0.9896}
    & \textbf{0.9886} & \textbf{0.9886} 
    & \underline{0.9465} & \underline{0.9453} 
    & \textbf{0.9942} & \textbf{0.9941}
    & \underline{0.9141} & 0.9141
    & \textbf{1} \\
    % Results
    \bottomrule
  \end{tabular}
  \label{tab:ISCXresults}
  \caption{Experimental Results on CIC-IoT, ISCX Tor-nonTor, and ISCX VPN-nonVPN Datasets w.r.t. \textit{Flow}-level Traffic Classification Task. (BOLD indicates the best score, and UNDERLINE denotes the second-best one. Avg. Rank indicates the average ranking of each model's results across all datasets and metrics.)}
\end{table*}
% Public Datasets

% Public Datasets
\begin{table*}[t]
    \centering
  \label{tab:ISCXresults_packet}
  \begin{tabular}{lccccccccccc}
    \toprule
    \multirow{2}{*}{Model} & \multicolumn{2}{c}{CIC-IoT} & \multicolumn{2}{c}{ISCX-Tor} & \multicolumn{2}{c}{ISCX-nonTor} & \multicolumn{2}{c}{ISCX-VPN} & \multicolumn{2}{c}{ISCX-nonVPN} & \multirow{2}{*}{\makecell{Avg.\\Rank}} \\
    \cmidrule(r){2-3} \cmidrule(r){4-5} \cmidrule(r){6-7} \cmidrule(r){8-9} \cmidrule(r){10-11}
    \multicolumn{1}{c}{} & AC & F1 & AC & F1 & AC & F1 & AC & F1 & AC & F1 & \multicolumn{1}{c}{} \\
    \midrule
    % Results
    Securitas-C4.5
    & 0.8675 & 0.8571
    & \underline{0.9520} & \underline{0.9429}
    & 0.8916 & 0.8652
    & 0.7250 & 0.7250
    & 0.5833 & 0.6212
    & 4 \\
    Securitas-SVM
    & 0.6526 & 0.7098
    & 0.8229 & 0.8046
    & 0.7333 & 0.7180
    & 0.6375 & 0.6454
    & 0.5750 & 0.6933
    & 9 \\
    Securitas-Bayes
    & 0.6667 & 0.7541
    & 0.8083 & 0.6938
    & 0.8062 & 0.7842
    & 0.6125 & 0.6736
    & 0.5416 & 0.6938
    & 10 \\
    2D-CNN
    & 0.7745 & 0.7863
    & 0.3781 & 0.3576
    & 0.6560 & 0.6877
    & 0.2887 & 0.2219
    & 0.5517 & 0.4638
    & 11 \\
    3D-CNN
    & 0.8788 & 0.8779
    & 0.7837 & 0.7534
    & 0.4791 & 0.4273
    & 0.8445 & 0.8436
    & 0.5346 & 0.5114
    & 8 \\
    DP-SAE
    & 0.7284 & 0.6134
    & 0.8190 & 0.8078
    & 0.8244 & 0.7889
    & 0.7227 & 0.6849
    & 0.6985 & 0.6952
    & 6 \\
    DP-CNN
    & 0.8664 & 0.8607
    & 0.6622 & 0.6342
    & 0.8784 & 0.8633
    & 0.9270 & 0.9283
    & 0.4326 & 0.3201
    & 7 \\
    BLJAN
    & 0.9643 & 0.9620
    & 0.7580 & 0.7762
    & 0.1348 & 0.2081
    & 0.8539 & 0.7646
    & 0.7356 & 0.7640
    & 5 \\
    EBSNN-GRU
    & 0.9440 & 0.9431
    & 0.9319 & 0.9208
    & \underline{0.9271} & \underline{0.9207}
    & 0.9467 & 0.9463
    & 0.8103 & 0.8096
    & 3 \\
    EBSNN-LSTM
    & \underline{0.9759} & \textbf{0.9819}
    & 0.7663 & 0.7183
    & \textbf{0.9519} & \textbf{0.9484}
    & \underline{0.9527} & \underline{0.9531}
    & \underline{0.8156} & \underline{0.8128}
    & 2 \\
    \midrule
    \textbf{MH-Net}
    & \textbf{0.9806} & \underline{0.9800}
    & \textbf{0.9916} & \textbf{0.9917}
    & 0.9156 & 0.9122
    & \textbf{0.9768} & \textbf{0.9766}
    & \textbf{0.8822} & \textbf{0.8814}
    & \textbf{1} \\
    % Results
    \bottomrule
  \end{tabular}
  \caption{Experimental Results on CIC-IoT, ISCX Tor-nonTor, and ISCX VPN-nonVPN Datasets w.r.t. \textit{Packet}-level Traffic Classification Task. (BOLD indicates the best score, and UNDERLINE denotes the second-best one. Avg. Rank indicates the average ranking of each model's results across all datasets and metrics.)}
\end{table*}
% Public Datasets

The comparison results of the flow-level and packet-level traffic classification tasks on the CIC-IoT and ISCX datasets are shown in Tables 1 and 2, respectively. 

\textbf{Flow-level Traffic Classification Results}. 
According to Table 1, we can draw several conclusions w.r.t. the flow-level task. 
(1) MH-Net achieves the overall best results w.r.t. all the metrics on the CIC-IoT and ISCX datasets, followed by TFE-GNN and YaTC, respectively. 
(2) Compared with traditional statistical feature methods, our approach surpasses them by a large margin due to the sufficient utilization of traffic bytes instead of statistical features. 
As for deep learning methods, MH-Net still has obvious advantages. 
(3) Among these methods, although TFE-GNN and YaTC also leverage raw bytes for representation learning, the overall performance is still significantly worse than that of MH-Net, which implies insufficient byte utilization and ignorance of fine-grained byte correlations in their model design. 
(4) Additionally, ET-BERT, which features a large number of parameters and pretraining on large-scale datasets, reaches decent results on the ISCX-nonVPN dataset, but its computational overhead is outrageously large.

\textbf{Packet-level Traffic Classification Results}. 
From Table 2, we can also conclude several observations and findings regarding the packet-level task. 
(1) MH-Net still has absolute superiority over other baselines, followed by EBSNN-LSTM and EBSNN-GRU. 
(2) Although EBSNN shows competitive results across all the datasets, its overall results are still inferior to MH-Net by a noticeable margin, which can be attributed to the fact that the byte segment in EBSNN cannot fully exploit the informative correlation between bytes, thus causing performance bottleneck. 
(3) Securitas, which uncovers n-gram keywords for frequency matching and ranks fourth according to the Avg. Rank in Table 2, performs much worse than MH-Net due to the inflexible and rigid keyword patterns, reflecting the importance and effectiveness of mining informative byte correlations.

From the overall perspective of the flow-level and packet-level tasks, MH-Net reaches the best comprehensive performance, verifying the effectiveness of our proposed model.

\subsection{Ablation Study (RQ2)}
\label{sec:exp_ablation}

For a clear understanding of MH-Net's architecture design, we conduct a comprehensive ablation study on the CIC-IoT and ISCX-VPN dataset w.r.t. F1-Score, and the results are shown in Table 3. 
From Table 3, we can conclude that the 8-bit traffic unit contributes more to the model performance than the 4-bit one. Still, the information carried by the latter cannot be ignored (the detailed analysis of traffic units will be presented later in RQ3). 
After we convert the heterogeneous traffic graph into a homogeneous one (i.e., with only one edge type), we can see a significant drop in the results, showing that it is necessary to model the heterogeneous correlations between traffic units (or bytes). 
We also investigate how contrastive learning tasks contribute to the results. 
The results indicate that integrating packet-level and flow-level contrastive loss can improve the classification performance of both tasks, with the latter having a more salient impact.

In short, MH-Net achieves the best results on flow-level and packet-level tasks compared to various variants.

% Ablation Table
\begin{table}[h]
    \centering
  \label{tab:Ablation_study}
  \begin{tabular}{lcccc}
    \toprule
    \multirow{2}{*}{Variant} & \multicolumn{2}{c}{CIC-IoT} & \multicolumn{2}{c}{ISCX-VPN} \\
    \cmidrule(r){2-3} \cmidrule(r){4-5}
    \multicolumn{1}{c}{} & Flow & Packet & Flow & Packet \\
    \midrule
    % Results
    w/o 4-bit View
    & 0.9800 & 0.9681 
    & 0.9498 & 0.9355 \\
    w/o 8-bit View
    & 0.8842 & 0.8970 
    & 0.1702 & 0.1488 \\
    w/o Hetero.
    & 0.9154 & 0.9551 
    & 0.9120 & 0.9010 \\
    w/o $L_\text{PCL}$
    & 0.9835 & 0.9618 
    & 0.9524 & 0.9524 \\
    w/o $L_\text{FCL}$
    & 0.9767 & 0.9631 
    & 0.9467 & 0.9204 \\
    \midrule
    \textbf{MH-Net}
    & \textbf{0.9896} & \textbf{0.9800} 
    & \textbf{0.9941} & \textbf{0.9766} \\
    % Results
    \bottomrule
  \end{tabular}
  \caption{Ablation Study of MH-Net on the CIC-IoT and ISCX-VPN Dataset w.r.t. F1-Score.}
\end{table}

\subsection{More Analysis on MH-Net (RQ3)}
\label{sec:exp_granularity}

In this section, we first analyze the sensitivity of MH-Net. Then, to investigate the impact of traffic units, we further analyze the choice and combination of traffic units.

\subsubsection{Sensitivity Analysis.} We conduct sensitivity analysis on the ISCX-VPN dataset to investigate the impact of the packet-level and flow-level contrastive loss ratios $\alpha$ and $\beta$. Figure 2 shows that (1) both tasks show a gradual improvement trend with the increase of $\alpha$, which implies the effectiveness of packet-level contrastive learning. 
(2) In contrast, the impact of $\beta$ on model performance fluctuates slightly and reaches its best at about $\beta=0.5$, which may be due to the excessive randomness caused by the augmentation of the traffic flow. One promising improvement may be introducing learnable augmentation to drop packets adaptively.

% Sensitivity Table
\begin{figure}[h]
    \centering
	\begin{subfigure}{0.49\linewidth}
		\centering
		\includegraphics[width=1.0\linewidth]{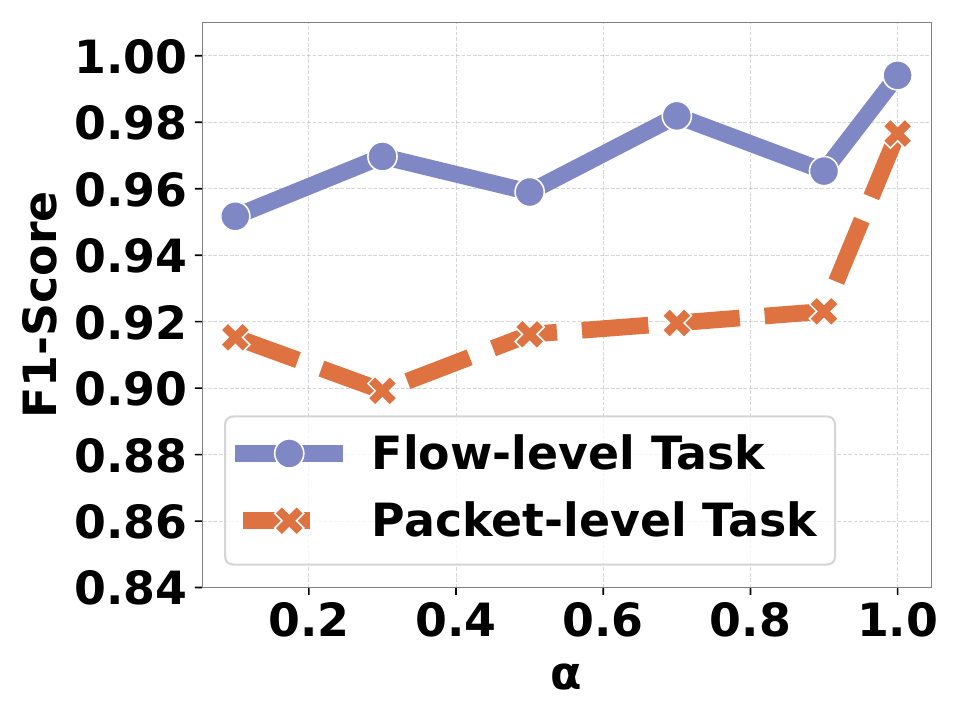}
	\end{subfigure}
	\centering
	\begin{subfigure}{0.49\linewidth}
		\centering
		\includegraphics[width=1.0\linewidth]{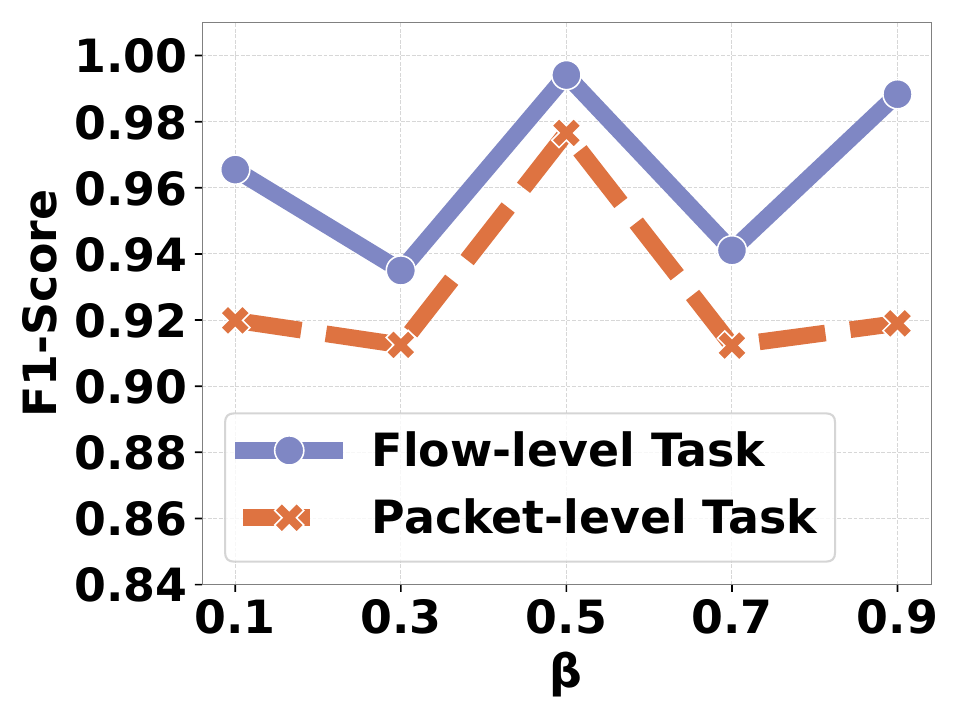}
	\end{subfigure}
	\caption{Sensitivity Analysis w.r.t. $\alpha$ and $\beta$ on the ISCX-VPN Dataset.}
	\label{sensitivity}
\end{figure}

\subsubsection{The Choice of Traffic Units.} As shown in Figure 3, we aggregate different numbers of traffic bits into traffic units and conduct experiments using single-view heterogeneous traffic graphs. Figure 3 shows that the 8-bit traffic unit achieves the best result, while others are much worse than it. 
In addition, both the flow-level and packet-level tasks show almost the same fluctuation trend w.r.t. F1-Score with traffic unit changes. 
The results also suggest that byte (i.e., 8-bit) is the primary unit of information transmission, and the reason for the decline in the performance of other traffic units may come from two aspects: 
(1) they break the integrity of bytes; (2) the traffic graph's size is limited by the traffic unit's bit length, which may be one of the influencing factors.

\begin{figure}[h]
	\centering
	\includegraphics[width=0.9\linewidth]{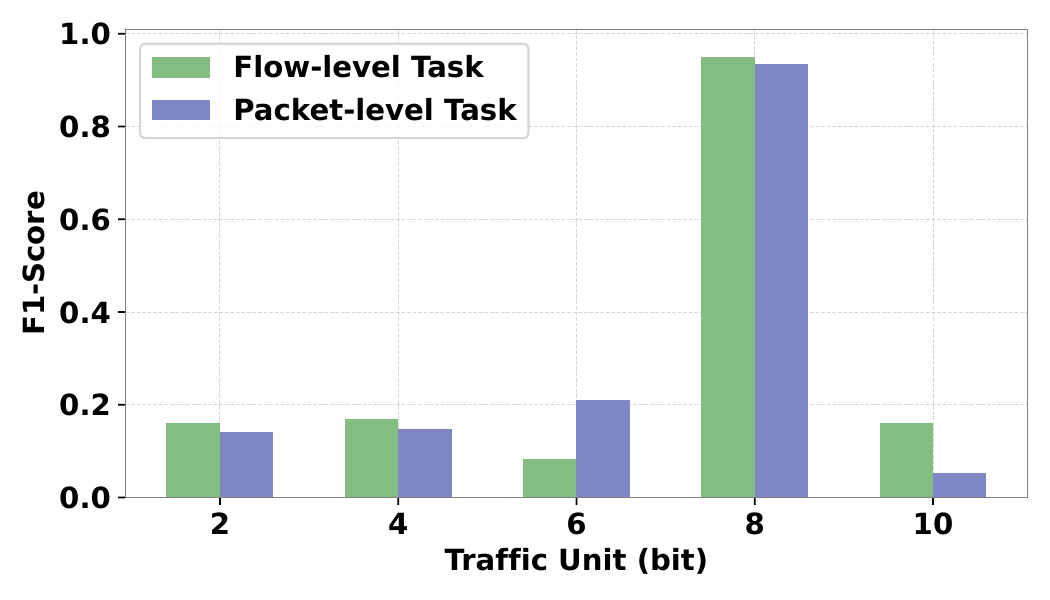}
        \caption{The Choice of Traffic Units on the ISCX-VPN Dataset.}
	\label{impact_unit}
\end{figure}

\subsubsection{The Combination of Traffic Units.} To further reveal how the combination of traffic units affects model performance, we conduct experiments on any two traffic unit combinations of 2, 4, 6, 8, and 10, and the result is shown in Figure 4. From Figure 4, we can draw the following conclusions. (1) The combination of 4\&8-bit traffic units reaches the best result, followed by the combination of 8\&10-bit traffic units. 
(2) Combining two traffic units can improve the model performance to a certain extent compared to a single traffic unit (e.g., 4\&6-bit, 4\&8-bit, 4\&10-bit, and 8\&10-bit traffic units), but it may also produce more negative results (e.g., 2\&8-bit and 6\&8-bit traffic units). 
This implies that there may be a trade-off between information complementarity and interference between traffic units with different information granularity, which can be potentially utilized to further improve model performance.

\begin{figure}[h]
	\centering
	\includegraphics[width=0.9\linewidth]{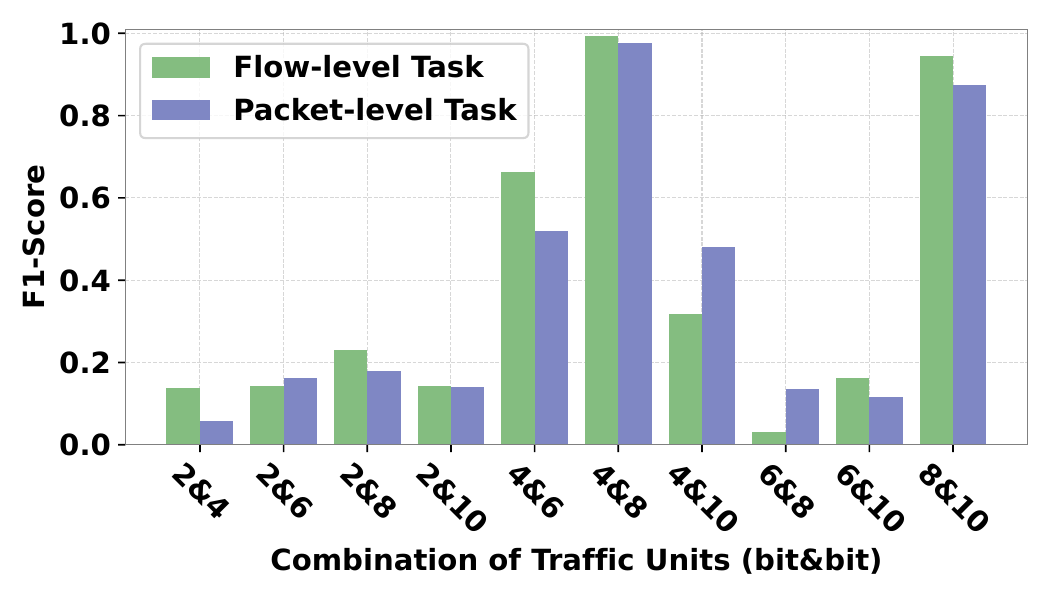}
        \caption{The Combination of Traffic Units on the ISCX-VPN Dataset.}
	\label{impact_comb}
\end{figure}

\section{Conclusion}
\label{sec:conclusion}
This paper proposes an effective model named MH-Net, which constructs multi-view heterogeneous traffic graphs via point-wise mutual information by aggregating different numbers of traffic bits into traffic units. 
In particular, we uncover the heterogeneous byte correlations contained in the traffic graph and employ a heterogeneous graph neural network for graph representation learning. 
Furthermore, we conduct supervised contrastive learning in a multi-task manner to obtain more robust traffic representations. 
The experimental results show that MH-Net reaches the overall best performance on both the flow-level and packet-level traffic classification tasks compared with dozens of baselines. In addition, an extensive analysis of traffic units reveals the possibility of using complementary information from different traffic units to improve traffic classification performance.

\section{Acknowledgments}

This work was supported in part by the Major Key Project of PCL (No. PCL2023A06-4), Open Foundation of Key Laboratory of Cyberspace Security, Ministry of Education of China, Henan Key Laboratory of Network Cryptography  (No. KLCS20240304), Natural Science Research Startup Foundation of Recruiting Talents of Nanjing University of Posts and Telecommunications (No. NY223164, NY224001), National Natural Science Foundation of China (No. 62022024), CCF-Huawei Populus Grove Fund, Jiangsu Provincial Key Laboratory of Network and Information  Security (No. BM2003201), Key Laboratory of Computer Network and Information Integration of Ministry of Education of China (No. 93K-9), and Collaborative Innovation Center of Novel Software Technology and Industrialization.

\bibliography{aaai25}

\begin{thebibliography}{41}
\providecommand{\natexlab}[1]{#1}

\bibitem[{Chen et~al.(2020)Chen, Kornblith, Norouzi, and Hinton}]{SimCLR}
Chen, T.; Kornblith, S.; Norouzi, M.; and Hinton, G. 2020.
\newblock A Simple Framework for Contrastive Learning of Visual Representations.
\newblock In \emph{International Conference on Machine Learning}, 1597--1607.

\bibitem[{Conti et~al.(2015)Conti, Mancini, Spolaor, and Verde}]{Conti}
Conti, M.; Mancini, L.~V.; Spolaor, R.; and Verde, N.~V. 2015.
\newblock Analyzing Android Encrypted Network Traffic to Identify User Actions.
\newblock \emph{IEEE Transactions on Information Forensics and Security}, 11(1).

\bibitem[{Dadkhah et~al.(2022)Dadkhah, Mahdikhani, Danso, Zohourian, Truong, and Ghorbani}]{ciciot}
Dadkhah, S.; Mahdikhani, H.; Danso, P.~K.; Zohourian, A.; Truong, K.~A.; and Ghorbani, A.~A. 2022.
\newblock Towards the development of a realistic multidimensional IoT profiling dataset.
\newblock In \emph{2022 19th Annual International Conference on Privacy, Security \& Trust (PST)}, 1--11. IEEE.

\bibitem[{Gil et~al.(2016)Gil, Lashkari, Mamun, and Ghorbani}]{ISCX}
Gil, G.~D.; Lashkari, A.~H.; Mamun, M.; and Ghorbani, A.~A. 2016.
\newblock Characterization of Encrypted and VPN Traffic Using Time-Related Features.
\newblock In \emph{International Conference on Information Systems Security and Privacy}, 407--414.

\bibitem[{Hamilton, Ying, and Leskovec(2017)}]{GraphSAGE}
Hamilton, W.; Ying, Z.; and Leskovec, J. 2017.
\newblock Inductive Representation Learning on Large Graphs.
\newblock In \emph{Conference on Neural Information Processing Systems}.

\bibitem[{Hayes and Danezis(2016)}]{KFP}
Hayes, J.; and Danezis, G. 2016.
\newblock k-fingerprinting: a Robust Scalable Website Fingerprinting Technique.
\newblock In \emph{USENIX Security Symposium}, 1187--1203.

\bibitem[{He et~al.(2020)He, Fan, Wu, Xie, and Girshick}]{MoCo}
He, K.; Fan, H.; Wu, Y.; Xie, S.; and Girshick, R. 2020.
\newblock Momentum Contrast for Unsupervised Visual Representation Learning.
\newblock In \emph{Computer Vision and Pattern Recognition}, 9729--9738.

\bibitem[{Hou et~al.(2022)Hou, Liu, Cen, Dong, Yang, Wang, and Tang}]{graphmae}
Hou, Z.; Liu, X.; Cen, Y.; Dong, Y.; Yang, H.; Wang, C.; and Tang, J. 2022.
\newblock Graphmae: Self-supervised masked graph autoencoders.
\newblock In \emph{Proceedings of the 28th ACM SIGKDD Conference on Knowledge Discovery and Data Mining}, 594--604.

\bibitem[{Khosla et~al.(2020)Khosla, Teterwak, Wang, Sarna, Tian, Isola, Maschinot, Liu, and Krishnan}]{SCL}
Khosla, P.; Teterwak, P.; Wang, C.; Sarna, A.; Tian, Y.; Isola, P.; Maschinot, A.; Liu, C.; and Krishnan, D. 2020.
\newblock Supervised Contrastive Learning.
\newblock In \emph{Conference on Neural Information Processing Systems}, 18661--18673.

\bibitem[{Lashkari et~al.(2017)Lashkari, Draper-Gil, Mamun, and Ghorbani}]{ISCX-Tor}
Lashkari, A.~H.; Draper-Gil, G.; Mamun, M. S.~I.; and Ghorbani, A.~A. 2017.
\newblock Characterization of Tor Traffic Using Time Based Features.
\newblock In \emph{International Conference on Information System Security and Privacy}, 253--262.

\bibitem[{Le-Khac, Healy, and Smeaton(2020)}]{CLSurvey}
Le-Khac, P.~H.; Healy, G.; and Smeaton, A.~F. 2020.
\newblock Contrastive Representation Learning: A Framework and Review.
\newblock \emph{IEEE Access}, 8: 193907--193934.

\bibitem[{Lim et~al.(2019)Lim, Kim, Heo, Kim, Hong, and Han}]{2DCNN}
Lim, H.-K.; Kim, J.-B.; Heo, J.-S.; Kim, K.; Hong, Y.-G.; and Han, Y.-H. 2019.
\newblock Packet-based Network Traffic Classification Using Deep Learning.
\newblock In \emph{International Conference on Artificial Intelligence in Information and Communication}, 046–051.

\bibitem[{Lin et~al.(2022)Lin, Xiong, Gou, Li, Shi, and Yu}]{ETBERT}
Lin, X.; Xiong, G.; Gou, G.; Li, Z.; Shi, J.; and Yu, J. 2022.
\newblock ET-BERT: A Contextualized Datagram Representation with Pre-training Transformers for Encrypted Traffic Classification.
\newblock In \emph{The Web Conference}, 633--642.

\bibitem[{Liu et~al.(2019)Liu, He, Xiong, Cao, and Li}]{FSNet}
Liu, C.; He, L.; Xiong, G.; Cao, Z.; and Li, Z. 2019.
\newblock FS-Net: A Flow Sequence Network For Encrypted Traffic Classification.
\newblock In \emph{IEEE Conference on Computer Communications}, 1171--1179.

\bibitem[{Lotfollahi et~al.(2020)Lotfollahi, Siavoshani, Zade, and Saberian}]{DeepPacket}
Lotfollahi, M.; Siavoshani, M.~J.; Zade, R. S.~H.; and Saberian, M. 2020.
\newblock Deep packet: a novel approach for encrypted traffic classification using deep learning.
\newblock \emph{Soft Computing}, 24(3): 1999--2012.

\bibitem[{Mao et~al.(2021)Mao, Xiao, Hu, Luo, Zhang, and Xia}]{BLJAN}
Mao, K.; Xiao, X.; Hu, G.; Luo, X.; Zhang, B.; and Xia, S. 2021.
\newblock Byte-Label Joint Attention Learning for Packet-grained Network Traffic Classification.
\newblock In \emph{International Workshop on Quality of Service}, 1--10.

\bibitem[{Meng et~al.(2022)Meng, Wang, Ma, Luo, Li, and Zhang}]{PacRep}
Meng, X.; Wang, Y.; Ma, R.; Luo, H.; Li, X.; and Zhang, Y. 2022.
\newblock Packet Representation Learning for Traffic Classification.
\newblock In \emph{ACM SIGKDD Conference on Knowledge Discovery and Data Mining}, 3546–3554.

\bibitem[{Panchenko et~al.(2016)Panchenko, Lanze, Zinnen, Henze, Pennekamp, Wehrle, and Engel}]{CUMUL}
Panchenko, A.; Lanze, F.; Zinnen, A.; Henze, M.; Pennekamp, J.; Wehrle, K.; and Engel, T. 2016.
\newblock Website Fingerprinting at Internet Scale.
\newblock In \emph{Annual Network and Distributed System Security Symposium}.

\bibitem[{Papadogiannaki and Ioannidis(2021)}]{EncrySurvey}
Papadogiannaki, E.; and Ioannidis, S. 2021.
\newblock A Survey on Encrypted Network Traffic Analysis Applications, Techniques, and Countermeasures.
\newblock \emph{ACM Computing Surveys}, 54(6): 1--35.

\bibitem[{Qiu et~al.(2020)Qiu, Chen, Dong, Zhang, Yang, Ding, Wang, and Tang}]{GCC}
Qiu, J.; Chen, Q.; Dong, Y.; Zhang, J.; Yang, H.; Ding, M.; Wang, K.; and Tang, J. 2020.
\newblock GCC: Graph Contrastive Coding for Graph Neural Network Pre-Training.
\newblock In \emph{ACM SIGKDD International Conference on Knowledge Discovery and Data Mining}, 1150–1160.

\bibitem[{Ramadhani(2018)}]{VPNTor}
Ramadhani, E. 2018.
\newblock Anonymity communication VPN and Tor: a comparative study.
\newblock \emph{Journal of Physics: Conference Series}, 983(1): 012060.

\bibitem[{Ren, Dubois, and Choffnes(2019)}]{Cross-plateform}
Ren, J.; Dubois, D.; and Choffnes, D. 2019.
\newblock An international view of privacy risks for mobile apps.

\bibitem[{Schroff, Kalenichenko, and Philbin(2015)}]{tripletloss}
Schroff, F.; Kalenichenko, D.; and Philbin, J. 2015.
\newblock FaceNet: A Unified Embedding for Face Recognition and Clustering.
\newblock In \emph{Computer Vision and Pattern Recognition}, 815--823.

\bibitem[{Shapira and Shavitt(2021)}]{FlowPic}
Shapira, T.; and Shavitt, Y. 2021.
\newblock FlowPic: A Generic Representation for Encrypted Traffic Classification and Applications Identification.
\newblock \emph{IEEE Transactions on Network and Service Management}, 18(2): 1218--1232.

\bibitem[{Sharma, Dangi, and Mishra(2021)}]{Encryption}
Sharma, R.; Dangi, S.; and Mishra, P. 2021.
\newblock A Comprehensive Review on Encryption based Open Source Cyber Security Tools.
\newblock In \emph{IEEE International Conference on Signal Processing, Computing and Control}, 614--619.

\bibitem[{Shen et~al.(2021)Shen, Zhang, Zhu, Xu, and Du}]{GraphDApp}
Shen, M.; Zhang, J.; Zhu, L.; Xu, K.; and Du, X. 2021.
\newblock Accurate Decentralized Application Identification via Encrypted Traffic Analysis Using Graph Neural Networks.
\newblock \emph{IEEE Transactions on Information Forensics and Security}, 16(1): 2367--2380.

\bibitem[{Sirinam et~al.(2018)Sirinam, Imani, Juarez, and Wright}]{DF}
Sirinam, P.; Imani, M.; Juarez, M.; and Wright, M.~K. 2018.
\newblock Deep Fingerprinting: Undermining Website Fingerprinting Defenses with Deep Learning.
\newblock In \emph{Conference on Computer and Communications Security}, 1928–1943.

\bibitem[{Taylor et~al.(2016)Taylor, Spolaor, Conti, and Martinovic}]{AppScanner}
Taylor, V.~F.; Spolaor, R.; Conti, M.; and Martinovic, I. 2016.
\newblock AppScanner: Automatic Fingerprinting of Smartphone Apps From Encrypted Network Traffic.
\newblock In \emph{IEEE European Symposium on Security and Privacy}, 439--454.

\bibitem[{Tong, Faloutsos, and Pan(2006)}]{random_walk}
Tong, H.; Faloutsos, C.; and Pan, J.-Y. 2006.
\newblock Fast random walk with restart and its applications.
\newblock In \emph{Sixth international conference on data mining (ICDM'06)}, 613--622. IEEE.

\bibitem[{van~den Oord, Li, and Vinyals(2018)}]{CPC}
van~den Oord, A.; Li, Y.; and Vinyals, O. 2018.
\newblock Representation Learning with Contrastive Predictive Coding.
\newblock \emph{arXiv preprint arXiv:1807.03748v2}.

\bibitem[{van Ede et~al.(2020)van Ede, Bortolameotti, Continella, Ren, Dubois, and Lindorfer}]{FlowPrint}
van Ede, T.; Bortolameotti, R.; Continella, A.; Ren, J.; Dubois, D.~J.; and Lindorfer, M. 2020.
\newblock FlowPrint: Semi-Supervised Mobile-App Fingerprinting on Encrypted Network Traffic.
\newblock In \emph{Annual Network and Distributed System Security Symposium}.

\bibitem[{Xiao et~al.(2022)Xiao, Xiao, Li, Luo, Zheng, and Xia}]{EBSNN}
Xiao, X.; Xiao, W.; Li, R.; Luo, X.; Zheng, H.; and Xia, S. 2022.
\newblock EBSNN: Extended Byte Segment Neural Network for Network Traffic Classification.
\newblock \emph{IEEE Transactions on Dependable and Secure Computing}, 19(5): 3521--3538.

\bibitem[{Xiao et~al.(2024)Xiao, Zhou, Yang, Yu, Zhang, Liu, and Luo}]{xiao2024comprehensive}
Xiao, X.; Zhou, X.; Yang, Z.; Yu, L.; Zhang, B.; Liu, Q.; and Luo, X. 2024.
\newblock A comprehensive analysis of website fingerprinting defenses on Tor.
\newblock \emph{Computers \& Security}, 136: 103577.

\bibitem[{Xu, Geng, and Jin(2022)}]{ETC-PS}
Xu, S.-J.; Geng, G.-G.; and Jin, X.-B. 2022.
\newblock Seeing Traffic Paths: Encrypted Traffic Classification With Path Signature Features.
\newblock \emph{IEEE Transactions on Information Forensics and Security}, 17(1): 2166--2181.

\bibitem[{Yao, Mao, and Luo(2019)}]{PMI}
Yao, L.; Mao, C.; and Luo, Y. 2019.
\newblock Graph Convolutional Networks for Text Classification.
\newblock In \emph{AAAI Conference on Artificial Intelligence}, 7370--7377.

\bibitem[{You et~al.(2020)You, Chen, Sui, Chen, Wang, and Shen}]{GraphCL}
You, Y.; Chen, T.; Sui, Y.; Chen, T.; Wang, Z.; and Shen, Y. 2020.
\newblock Graph Contrastive Learning with Augmentations.
\newblock In \emph{Conference on Neural Information Processing Systems}, 5812--5823.

\bibitem[{Yun et~al.(2015)Yun, Wang, Zhang, and Zhou}]{Securitas}
Yun, X.; Wang, Y.; Zhang, Y.; and Zhou, Y. 2015.
\newblock A Semantics-Aware Approach to the Automated Network Protocol Identification.
\newblock \emph{IEEE/ACM Transactions on Networking}, 24(1): 583--595.

\bibitem[{Zhang et~al.(2023)Zhang, Yu, Xiao, Li, Mercaldo, Luo, and Liu}]{TFE-GNN}
Zhang, H.; Yu, L.; Xiao, X.; Li, Q.; Mercaldo, F.; Luo, X.; and Liu, Q. 2023.
\newblock TFE-GNN: A Temporal Fusion Encoder Using Graph Neural Networks for Fine-grained Encrypted Traffic Classification.
\newblock In \emph{The Web Conference}, 2066–2075.

\bibitem[{Zhang et~al.(2020)Zhang, Li, Ye, and Wu}]{3DCNN}
Zhang, J.; Li, F.; Ye, F.; and Wu, H. 2020.
\newblock Autonomous Unknown-Application Filtering and Labeling for DL-based Traffic Classifier Update.
\newblock In \emph{IEEE Conference on Computer Communications}, 397–405.

\bibitem[{Zhao et~al.(2023)Zhao, Zhan, Deng, Wang, Wang, Gui, and Xue}]{YaTC}
Zhao, R.; Zhan, M.; Deng, X.; Wang, Y.; Wang, Y.; Gui, G.; and Xue, Z. 2023.
\newblock Yet another traffic classifier: A masked autoencoder based traffic transformer with multi-level flow representation.
\newblock In \emph{Proceedings of the AAAI Conference on Artificial Intelligence}, volume~37, 5420--5427.

\bibitem[{Zheng et~al.(2020)Zheng, Gou, Yan, and Mo}]{RBRN}
Zheng, W.; Gou, C.; Yan, L.; and Mo, S. 2020.
\newblock Learning to Classify: A Flow-Based Relation Network for Encrypted Traffic Classification.
\newblock In \emph{The Web Conference}, 13--22.

\end{thebibliography}

\clearpage
\appendix
\section{Threat Model and Assumptions}
\label{sec:appendix_threat}

We briefly describe the threat model and assumptions as follows. 

Normal users employ mobile apps to communicate with remote servers. The attacker is a passive observer (i.e., he cannot decrypt or modify packets). The attacker captures the packets of the target apps by compromising the device or sniffing the network link. Then, the attacker analyzes the captured packets to infer the behaviors of normal users.

\section{More Experimental Details}
\label{sec:appendix_exp_detail}

\subsection{Dataset Details}

We adopt the CIC-IoT, ISCX VPN-nonVPN, and ISCX Tor-nonTor in this work. 

The CIC-IoT~\cite{ciciot} dataset consists of network traffic of IoT devices. 
The ISCX VPN-nonVPN~\cite{ISCX} dataset is a collection of two datasets: the ISCX-VPN and ISCX-nonVPN datasets. The ISCX-VPN dataset contains traffic collected over virtual private networks (VPNs), which are commonly used for accessing blocked websites or services. Due to obfuscation technology, this kind of traffic can be challenging to detect. In contrast, the ISCX-nonVPN dataset contains regular traffic not collected over VPNs. 
The ISCX Tor-nonTor~\cite{ISCX-Tor} dataset comprises the ISCX-Tor and ISCX-nonTor datasets. The ISCX-Tor dataset involves traffic collected over the onion router (Tor), making its traffic difficult to trace, whereas the ISCX-nonTor is a regular dataset not collected over Tor. 

In the experiment, we divide the traffic data in the CIC-IoT, ISCX VPN-nonVPN, and ISCX Tor-nonTor datasets into six, six, and eight categories, respectively, according to the type of traffic following the dataset description. The dataset statistics are presented in Table 1. 

\begin{table}[h]
  \centering
  \label{tab:dataset_sta}
  \begin{tabular}{lccc}
    \toprule
    Dataset & \#Flow & \#Packet & \#Category \\
    \midrule
    % Results
    CIC-IoT
    & 2961 & 39208 & 6 \\
    ISCX-VPN
    & 1674 & 19282 & 6 \\
    ISCX-NonVPN
    & 3928 & 33838 & 6 \\
    ISCX-Tor
    & 1697 & 22888 & 8 \\
    ISCX-NonTor
    & 7979 & 68024 & 8 \\
    % Results
    \bottomrule
  \end{tabular}
    \caption{Dataset Statistics}
\end{table}

We give specific category divisions for each dataset in the following: 
\begin{itemize}
\item \textbf{CIC-IoT}: Power, Idle, Interactions, Scenarios, Active, Attacks. 
\item \textbf{ISCX-VPN}: VoIP, Streaming, P2P, File, Email, Chat. 
\item \textbf{ISCX-NonVPN}: VoIP, Video, Streaming, File, Email, Chat. 
\item \textbf{ISCX-Tor}: VoIP, Video, P2P, Mail, File, Chat, Browsing, Audio. 
\item \textbf{ISCX-NonTor}: VoIP, Video, P2P, FTP, Email, Chat, Browsing, Audio. 
\end{itemize}

As for category types of the ISCX VPN-NonVPN dataset, some files can be labeled as either “Browser” or other types like “Streaming.” So, we abandoned “Browser” and used the remaining six types. Note that we didn’t delete data but used the alternative types instead.

\subsection{Pre-processing}

For each dataset, we use SplitCap to obtain bidirectional flows from each pcap file. 
Due to the limited number of flows in the ISCX-Tor dataset, we enrich traffic flows by dividing each flow into 60-second non-overlapping blocks in our experiments~\cite{FlowPic}. 
Following~\cite{TFE-GNN}, we filter out traffic flows without payload or that surpass a length of 10000. Such flows usually establish connections between two communicating entities or result from temporary network failures, which have little useful information for classification. As for each packet in a traffic flow, we remove the Ethernet header, IP addresses, and port numbers to protect sensitive information while eliminating the potential interference it brings. After the above processing, we only keep the first 15 packets of a traffic flow at most to ensure relatively smaller computation costs, which is enough to achieve good performance.

Since our method performs both the flow-level and packet-level classification tasks, we first adopt stratified sampling to partition the flow-level training and testing dataset into 9:1 according to the number of traffic flows for all datasets. All packets in the flow-level training and testing datasets are directly used as the packet-level training and testing datasets, respectively. The category of each packet is consistent with the traffic flow to which it belongs. Note that independent packets can also be used to evaluate the packet-level task. Here, for convenience, the packets in a traffic flow are directly used for evaluation.

\section{Hyper-parameters}
\label{sec:appendix_hyper}

In this section, we give hyper-parameters used in this paper in Table 2.

% Hyper Table
\begin{table*}[t]
  \centering
  \label{tab:hyper-parameters}
  \begin{tabular}{lccccc}
    \toprule
    Datasets & ISCX-VPN & ISCX-NonVPN & ISCX-Tor & ISCX-NonTor & CIC-IoT \\
    \midrule
    % Results
    Batch Size
    & 16 & 102 & 32 & 102 & 24 \\
    Gradient Accumulation
    & 1 & 5 & 1 & 5 & 1\\
    Epoch
    & 20 & 120 & 100 & 120 & 20\\
    Learning Rate (Max, Min)
    & (1e-2, 1e-4) & (1e-2, 1e-5) & (1e-2, 1e-4) & (1e-2, 1e-4) & (1e-2, 1e-4)\\
    Label Smoothing
    & 0.0 & 0.01 & 0.0 & 0.0 & 0.0\\
    Warm Up
    & 0.1 & 0.1 & 0.1 & 0.1 & 0.1\\
    GNN Dropout Ratio
    & 0.0 & 0.1 & 0.0 & 0.2 & 0.0\\
    LSTM Dropout Ratio
    & 0.0 & 0.15 & 0.0 & 0.1 & 0.0 \\
    Embedding Dimension
    & 64 & 64 & 64 & 64 & 64\\
    Hidden Dimension
    & 128 & 128 & 128 & 128 & 128\\
    PMI Window Size
    & 5 & 5 & 5 & 5 & 5\\
    \midrule
    Packet Dropping Ratio $P_{\text{PD}}$
    & 0.6 & 0.6 & 0.6 & 0.6 & 0.6\\
    Temperature $\tau$
    & 0.07 & 0.07 & 0.07 & 0.07 & 0.07\\
    Flow-level Contrastive Loss Ratio $\beta$
    & 0.5 & 0.8 & 1.0 & 1.0 & 0.5\\
    Packet-level Contrastive Loss Ratio $\alpha$
    & 1.0 & 0.4 & 0.4 & 0.6 & 1.0\\
    % Results
    \bottomrule
  \end{tabular}
  \caption{Hyper-parameters}
\end{table*}

\section{Comparison with More Datasets}

In this section, we conduct extensive experiments on the CrossPlatform(Android)~\cite{Cross-plateform} dataset to further demonstrate the effectiveness of MH-Net, and the results are shown in Tables 3 and 4. 

From Tables 3 and 4, we can observe that MH-Net outperforms dozens of baselines and achieves SOTA performance on both the packet-level and flow-level classification tasks.

% Public Datasets
\begin{table}[t]
    \centering
  \begin{tabular}{lccc}
    \toprule
    \multirow{2}{*}{Model} & \multicolumn{2}{c}{CrossPlatform(Android)} & \multirow{2}{*}{\makecell{Avg.\\Rank}} \\
    \cmidrule(r){2-3}
    \multicolumn{1}{c}{} & AC & F1 & \multicolumn{1}{c}{} \\
    \midrule
    % Results
    AppScanner
    & 0.7812 & 0.7517
    & 6 \\
    K-FP
    & 0.8438 & 0.8218
    & 4 \\
    CUMUL
    & 0.7188 & 0.7092
    & 8 \\
    ETC-PS
    & 0.8125 & 0.7931
    & 5 \\
    \midrule
    FS-Net
    & 0.5625 & 0.4919
    & 9 \\
    DF
    & 0.4375 & 0.4417
    & 10 \\
    ET-BERT
    & \textbf{0.9893} & 0.9869
    & 2 \\
    GraphDApp
    & 0.7500 & 0.7255
    & 7 \\
    TFE-GNN
    & 0.9818 & 0.9818
    & 3 \\
    YaTC
    & \underline{0.9889} & \underline{0.9872}
    & 2 \\
    \midrule
    \textbf{MH-Net (Ours)}
    & \textbf{0.9893} & \textbf{0.9893}
    & \textbf{1} \\
    % Results
    \bottomrule
  \end{tabular}
  \caption{Experimental Results on the CrossPlatform(Android) Dataset w.r.t. \textbf{Flow}-level Traffic Classification Task. (\textbf{BOLD} indicates the best score, and \underline{UNDERLINE} denotes the second best one. Avg. Rank indicates the average ranking of each model's results across all datasets and metrics.)}
\end{table}
% Public Datasets

% Public Datasets
\begin{table}[t]
    \centering
  \begin{tabular}{lccc}
    \toprule
    \multirow{2}{*}{Model} & \multicolumn{2}{c}{CrossPlatform(Android)} & \multirow{2}{*}{\makecell{Avg.\\Rank}} \\
    \cmidrule(r){2-3}
    \multicolumn{1}{c}{} & AC & F1 & \multicolumn{1}{c}{} \\
    \midrule
    % Results
    Securitas-C4.5
    & 0.6444 & 0.6847
    & 4 \\
    Securitas-SVM
    & 0.5333 & 0.6331
    & 6 \\
    Securitas-Bayes
    & 0.5388 & 0.6891
    & 4 \\
    2D-CNN
    & 0.1602 & 0.0922
    & 10 \\
    3D-CNN
    & 0.4440 & 0.4007
    & 7 \\
    DP-SAE
    & 0.6608 & 0.5897
    & 5 \\
    DP-CNN
    & 0.2136 & 0.1514
    & 8 \\
    BLJAN
    & 0.1348 & 0.2081
    & 9 \\
    EBSNN-GRU
    & 0.9069 & 0.8998
    & 3 \\
    EBSNN-LSTM
    & \underline{0.9832} & \underline{0.9831}
    & 2 \\
    \midrule
    \textbf{MH-Net (Ours)}
    & \textbf{0.9875} & \textbf{0.9875}
    & \textbf{1} \\
    % Results
    \bottomrule
  \end{tabular}
  \caption{Experimental Results on the CrossPlatform(Android) Dataset w.r.t. \textbf{Packet}-level Traffic Classification Task. (\textbf{BOLD} indicates the best score, and \underline{UNDERLINE} denotes the second best one. Avg. Rank indicates the average ranking of each model's results across all datasets and metrics.)}
\end{table}
% Public Datasets

\end{document}